\documentclass[runningheads]{cl2emult}

\usepackage{makeidx}  
\usepackage{graphicx} 
\usepackage{subeqnar} 
\usepackage{multicol} 
\usepackage{cropmark} 
\usepackage{eso}      
\makeindex            

\newcommand{\euler}[1]{{\usefont{U}{eur}{m}{n}#1}}

\newcommand{\umu}{\mbox{\euler{\char22}}}

\begin{document}
\title*{SCUBA Deep Fields and Source Confusion}
%
%
\toctitle{SCUBA Deep Fields and Source Confusion} 
%
%
\titlerunning{SCUBA Deep Fields}
%
\author{Andrew~W.~Blain} 
\authorrunning{Andrew~W.~Blain}
%
%
\institute{Institute of Astronomy, Madingley Road, Cambridge, CB3 0HA, UK} 

\maketitle              

\begin{abstract}
Deep submillimetre(submm)-wave surveys made over the last three years using the 
SCUBA camera at the 15-m James Clerk Maxwell Telescope (JCMT) have revealed a 
new population of very luminous high-redshift galaxies. The properties of this 
population, and their contribution to the intensity of the extragalactic background 
radiation field are described briefly, especially in the context of the SCUBA lens
survey.\footnote{The SCUBA Lens Survey has been carried out since 1997 by 
Ian Smail, Rob Ivison, Jean-Paul Kneib and the author: a full description of the 
survey and references to supporting work can be found in the recent catalogue 
paper \cite{Smail_cat}} 
The potential problems caused by source confusion in the large 15-arcsec SCUBA 
observing beam for the identification and follow up of the results are discussed. 
The effects of confusion are not important for the study of 850-$\mu$m SCUBA 
sources brighter than about 2\,mJy. 
\end{abstract}

\begin{figure}[t]
\begin{center}
\includegraphics[width=.25\textwidth,angle=-90]{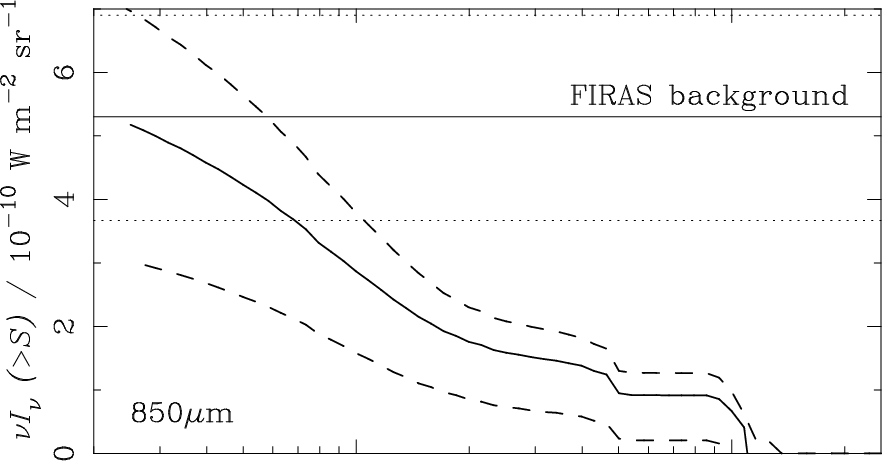} {\hspace{5pt}}
\includegraphics[width=.25\textwidth,angle=-90]{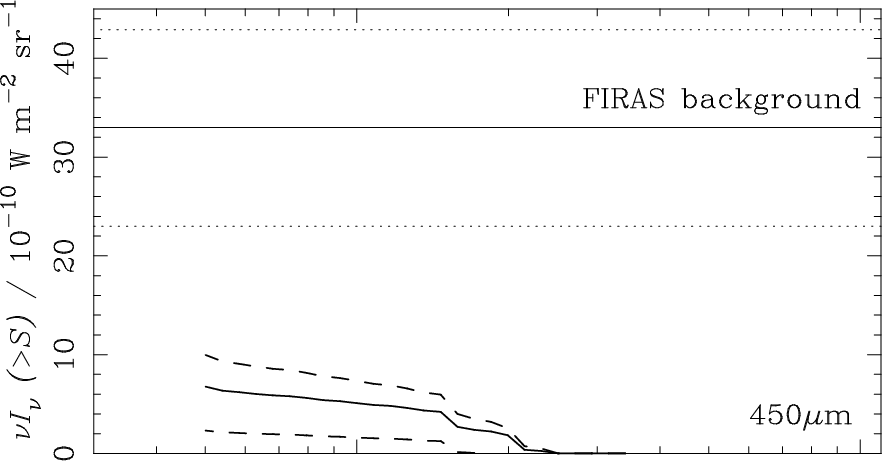}
\includegraphics[width=.5\textwidth,angle=-90]{blain_DF850.cps} {\hspace{5pt}}
\includegraphics[width=.5\textwidth,angle=-90]{blain_DF450.cps}
\end{center}
\caption[]{A compilation of published counts from SCUBA surveys at 850 and
450\,$\umu$m (lower panels), and the fraction of the background radiation 
intensity $\nu I_\nu$ \cite{Fixsen} contributed by sources exceeding a certain 
flux density $S$ in the SCUBA Lens Survey (upper panels)\cite{BIKS,BKIS}, as 
shown by the dotted lines in the lower panels. Atmospheric noise is more 
dominant in the 
SCUBA results at 450\,$\umu$m. The key to the references is; BKIS \cite{BKIS}, 
BIKS \cite{BIKS}, E00 \cite{E00}, C00 \cite{C00}, BCR \cite{BCR}, C01 \cite{C01}, 
Borys \cite{Borys}, Dunlop \cite{Dunlop}, Hu98 \cite{Hu98}, S97 \cite{SIB}, Ba98 
\cite{Ba98}, Ba99 \cite{Ba99}, E99 \cite{E99}.
}
\end{figure}

\subsubsection{High-redshift surveys for dusty galaxies} 

In 1997 the SCUBA camera at the JCMT \cite{Holland} became available, providing 
deep images of 5-arcmin$^2$ fields at 450 and 850\,$\umu$m, and allowing the first 
deep blank-field searches for redshifted thermal dust emission from distant 
galaxies. By exploiting the positive magnification bias and rich archival data in the 
fields of seven massive clusters of galaxies, 15 non-cluster sources and 2 cD galaxies 
\cite{Edge} were soon detected in the SCUBA Lens Survey \cite{SIB,Smail_cat} and 
several sources were identified \cite{Bargerspec,ERO}, three having certain 
identifications and redshifts based on mm-wave CO spectroscopy \cite{FI,FII,K}. 
Several other deep surveys in clusters \cite{C00,KKK} and non-cluster fields 
\cite{Ba99,Hu98,E00,Dunlop} have confirmed the existence of this population of 
distant galaxies. Three SCUBA-selected galaxies from all these surveys, without 
CO detections, have accurate positions from mm-wave continuum interferometry.  
\cite{Downes,FIII,Gear} Targeted SCUBA observations to image or obtain submm-wave 
photometry in the fields of both faint radio sources with very faint K-band counterparts 
\cite{BCR,C01} and known high-redshift objects \cite{Ivison_4C,CLBG,QSO,Arch} have 
increased the number of submm-selected galaxies known. In the first approach, the 
wide fields and accurate astrometry available from deep radio surveys is exploited to 
reveal relatively bright submm-selected galaxies \cite{BCR}. In the second, valuable 
redshifts are available for the detected submm sources. 

A summary of the published submm-wave counts at both 850 and 450\,$\umu$m is 
presented in Fig.\,1, alongside the contribution of the SCUBA Lens Survey galaxies to 
the integrated background radiation intensity derived from observations using 
{\it COBE} \cite{Fixsen}. At 850\,$\mu$m, a significant fraction 
of the background radiation has been associated with SCUBA sources brighter than
1\,mJy after correcting for the effects of lens magnification \cite{BKIS}. 

SCUBA surveys provide information about the form of evolution of high-redshift 
galaxies. The results indicate that a very luminous population of dust-enshrouded 
high-redshift galaxies exist, and that they emit a greater luminosity than optically 
selected high-redshift galaxies \cite{BSIK,BJ,Ringberg}. The detected galaxies are 
almost certainly still observed on the Rayleigh--Jeans side of their thermal dust 
SEDs, and so uncertainties in the dust temperature of the sources feed through into 
significant uncertainties in their total luminosity. However, by combining information 
from radio and 15-$\umu$m {\it ISO} mid-infrared(IR) flux densities of submm-selected 
sources, where available, from the extragalactic background spectrum at 
submm wavelengths, and from multiple-band mm/submm detections of objects with 
known redshifts, it is likely that a dust temperature of about 40\,K is typical 
\cite{BSIK,TBG,Hyfest,UMass}. This is thus a reasonable temperature to assume for the 
bulk of the population; if a greater value is assumed, then the luminosity density is 
likely to be overestimated. 

There has been a suggestion \cite{Lawrence} that a fraction of faint SCUBA sources 
could be very local, cold dust clouds in the Milky Way. In the light of the surface density 
of detectable SCUBA sources being increased in the fields of $z \simeq 0.2$ galaxy 
clusters -- presumably due to magnification bias -- this seems unlikely. In addition, the 
850-$\umu$m counts (Fig.\,1) are known to be steep between flux densities of 2 and 
10\,mJy, and yet a very local population of sources would have counts with a Euclidean 
slope or flatter. There are also no obvious signs of the brighter cousins of such objects 
in a 1800-deg$^2$ 400-GHz survey field \cite{BOOM}. Currently it seems unlikely 
that this suggestion is correct. 

A significant difficulty in studying SCUBA-selected galaxies is the relatively coarse 
15-arcsec resolution of 850-$\umu$m SCUBA images. The effect of redshifting the 
thermal dust emission spectrum of distant galaxies, which peaks at restframe far-IR 
wavelengths of order 60 to 100\,$\umu$m, into the submm waveband is to overcome 
the inverse square law, and leads to the standard approximately flat submm-wave flux 
density--redshift relation for $0.5 < z < 10$. The detection of a submm-wave source 
at a single wavelength thus provides almost no redshift information. The combination 
of these two effects -- coarse resolution and
potential high redshift -- makes the certain association of submm sources with 
counterparts in other wavebands very difficult, as recently discussed elsewhere 
\cite{Smail_UMass}. So far, extremely deep radio images have provided the best route 
for reliable follow-up of submm-selected galaxies \cite{SIOBK}, because of the 
apparently fairly universal radio-to-far-IR template SED for both nearby and distant 
galaxies combined with the large field and excellent sensitivity of the VLA \cite{Carilli}. 

\subsubsection{Source Confusion Noise}

There is a further difficulty for interpreting the results of these surveys -- the 
surface density of sources in the faintest images is as great as 0.06\,beam$^{-1}$ 
\cite{Hu98}, at which source confusion provides a significant contribution to the 
noise in the images. Confusion arises because of the uncertain and varying 
contribution of flux density from the numerous unresolved, undetected 
faint sources that fall within each observing beam. First recognized as 
a problem by early radio astronomers, confusion leads to a non-Gaussian
distribution of random intensity fluctuations on the sky, whose properties depend on 
the details of both the shape of the counts and the clustering strength of the galaxies in 
the survey. Confusion becomes the dominant source of noise for any observations deeper 
than a certain limit, which generally corresponds to a density of sources on the sky that  
is greater than about 0.03\,beam$^{-1}$. A recent summary of references to confusion 
and a series of simulations of its effects have been presented by Hogg \cite{Hogg}. 
Although a variety of analytic schemes have been suggested to study the effects of source 
confusion \cite{Toff}, numerical simulations are essential in order to make 
reliable statements, as long as they are based on accurate count models. The 
results of such simulations, specifically for the SCUBA observing parameters are shown 
in Fig.\,2. The flux density at which the surface density of sources 
exceeds 1\,beam$^{-1}$ generally provides a reasonable estimate of the width of the
distribution of confusion fluctuations: in Fig.\,2 it is lower by a factor of about two. 
The practical survey limit is an order of magnitude greater 
than the 1\,beam$^{-1}$ value, which corresponds to the traditional rule-of-thumb 
source density of 0.03\,beam$^{-1}$. The fluctuation distribution can be approximated 
reasonably by a log-normal distribution. 

\begin{figure}[t]
\begin{center}
\includegraphics[width=.53\textwidth,angle=-90]{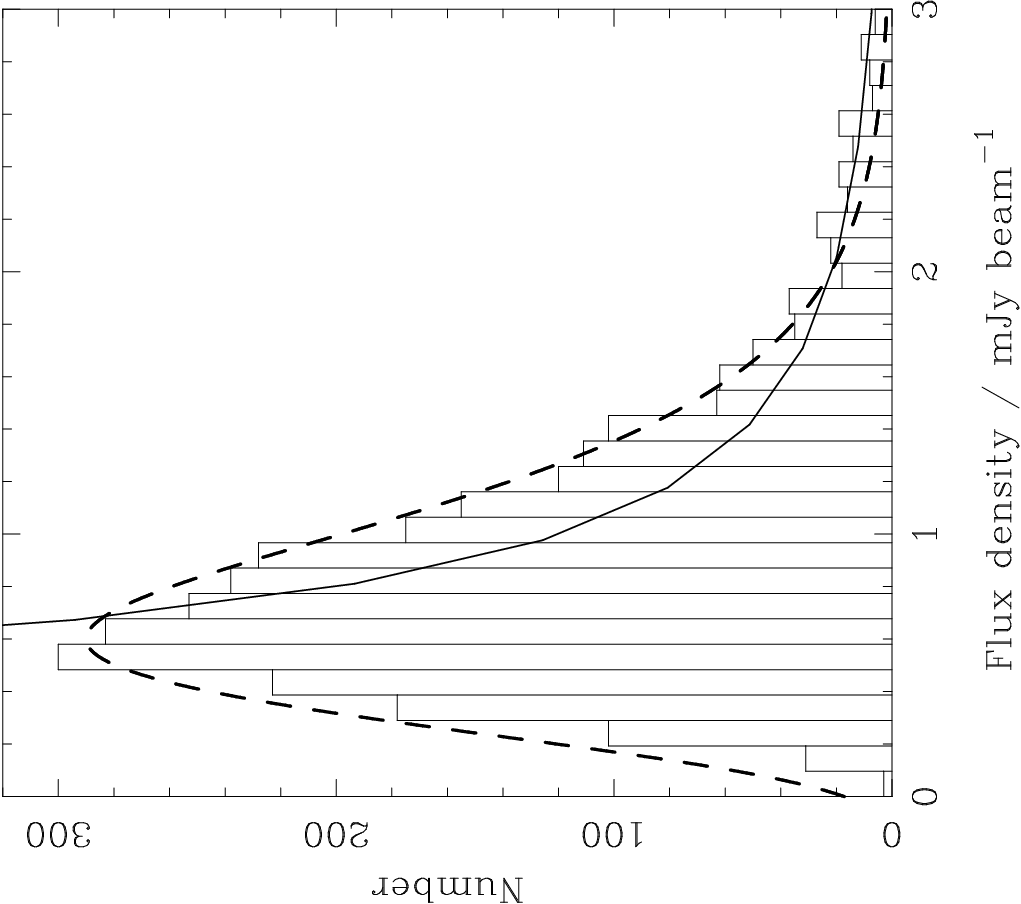} 
{\hspace{5pt}}
\includegraphics[width=.53\textwidth,angle=-90]{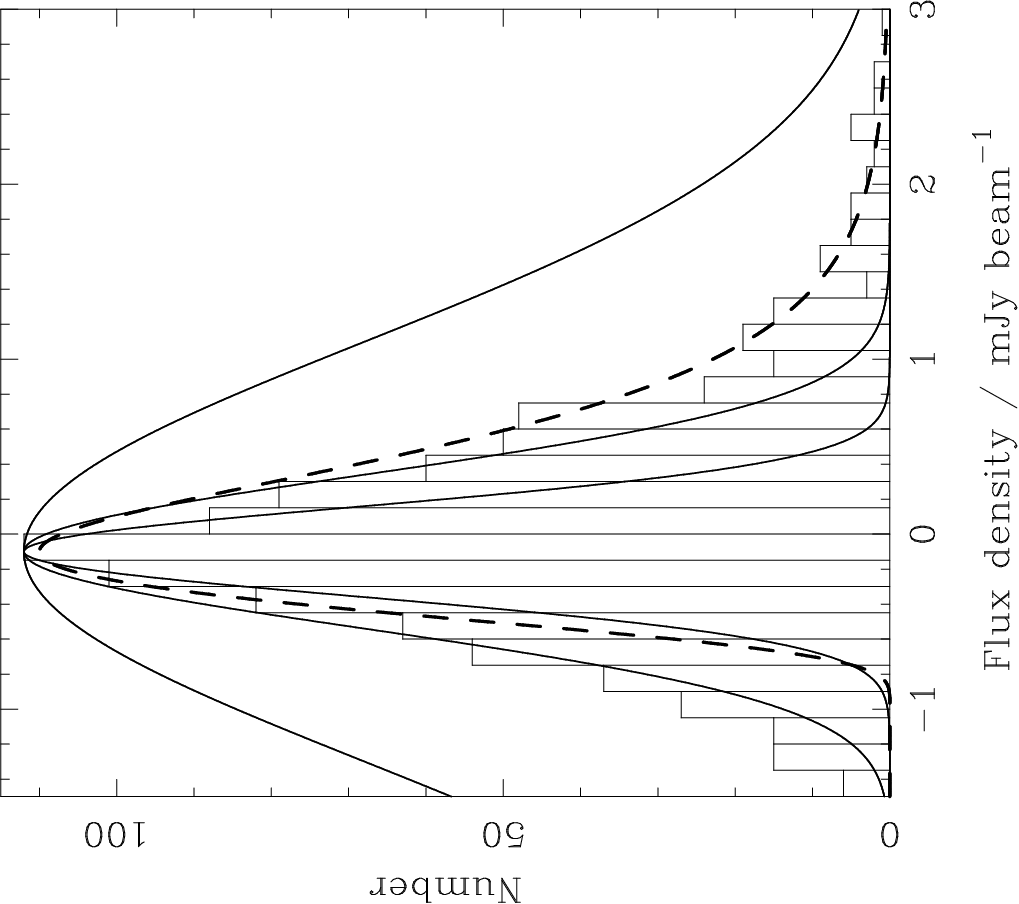}
\end{center}
\caption[]{The results of confusion simulations for the 15-arcsec SCUBA beam at 
850\,$\umu$m, based on a source population represented by the revised luminosity 
evolution model \cite{Ringberg,BSIK}. The histogram in the left panel represents the 
results of 3000 simulated observations of the direct distribution of sources on the 
sky. The underlying source count model is represented by the superimposed solid line. 
A log-normal distribution -- 
${\rm exp}\{-[{\rm ln}(S-\bar S+1)]^2/2[{\rm ln}(1+\sigma)]^2\}$ with $\bar S = 0.6$\,mJy 
and $\sigma = 0.47$\,mJy -- which represents the results accurately is shown by the 
dashed line. The histogram in the right panel represents the results of 1000 simulations 
of the sky observed using the SCUBA 3-beam chopping pattern. Three solid Gaussians 
are superimposed to represent sky and instrument noise levels of 0.25, 0.44 and 
1.5\,mJy\,beam$^{-1}$, which correspond to the simple 1\,beam$^{-1}$ confusion noise 
estimate, to the noise level in the HDF SCUBA image \cite{Hu98} and to the typical noise 
level in the SCUBA lens survey fields \cite{SIB,Smail_cat} respectively. The dashed line 
represents a log-normal distribution with $\bar S = -0.1$\,mJy and $\sigma = 0.52$\,mJy, 
which provides a reasonable description of the distribution at interesting positive flux
densities.}
\end{figure}

The deepest SCUBA surveys are as close to the confusion limit as any observations 
that have been interpreted to derive information about cosmic evolution, and so it is 
important to understand the effects of confusion. Since the first determination of the 
submm-wave counts \cite{SIB}, several studies of the effects of confusion noise in 
these surveys have been made \cite{BIS,Hogg,E00}, sometimes with different results. 
Hogg \cite{Hogg} emphasizes the importance of limiting the depth of a survey to 
avoid exceeding 0.03 sources per beam, and that the practical confusion limit is about 
an order of magnitude brighter than the approximate 1-$\sigma$ value quoted in the 
SCUBA paper \cite{BIS}. The effect is illustrated clearly in Fig.\,2: the 1-$\sigma$ estimate 
is only about 0.25\,mJy, but it is not safe to detect sources fainter than about 2\,mJy. This 
conclusion is supported by a study of the properties of correlated noise in the SCUBA 
image of the HDF \cite{Peacock}. Hogg also shows that the positions of submm-detected 
sources become impossible to determine accurately at a flux density brighter than the 
limit for accurate photometry. In contrast, Eales et al. \cite{E00} report a systematic 
increase in the flux density determined for a submm-selected galaxy due to the effects 
of confusion, which takes the form of a flux-density-independent scaling that is still 
significant for 10-mJy 850-$\umu$m sources. This disagrees with both the results 
of Hogg and those presented in Fig.\,2 here. Only SCUBA surveys with 850-$\umu$m 
detection sensitivities less than 2\,mJy are likely to suffer from the effects of confusion. 

Most of the well-studied SCUBA-selected galaxies have been drawn from the fields of 
lensing clusters of galaxies \cite{Smail_cat,KKK}. The magnification bias introduced by 
the clusters has the effect of reducing the contribution of undetected galaxies at a given 
flux density, and thus reducing the effect of confusion \cite{BSZ}. However, so far no 
observations have been made in cluster fields to a depth less than 1\,mJy, and so this 
effect has not been observed directly. A joint Netherlands--UK SCUBA project -- lead by 
John Peacock and Paul van der Werf -- is underway to obtain a very deep submm-wave 
image of the cluster A\,2218, and so probe the ultimate sensitivity of an 
850-$\umu$m camera on the 15-m JCMT. 

The results of the simulations shown in Fig.\,2 can readily be extended to address the 
importance of source confusion for other mm/submm and far-IR surveys. When we 
first derived estimates of confusion noise from early SCUBA surveys \cite{BIS}, much
less information was available about how to extrapolate to other beamsizes and 
observing frequencies. However, the availability of both additional data and the 
resulting revised models \cite{Ringberg} now provides the opportunity to 
make accurate predictions of the effects for other far-IR and mm/submm-wave 
telescopes.

\subsubsection{Conclusion} 

The results of a wide range of deep survey observations carried out using the SCUBA 
camera at the JCMT have been reviewed. The consequences of source confusion for the 
results has been discussed. SCUBA surveys in which the detection limit is shallower than 
about 2\,mJy are not likely to be affected significantly by confusion. 

\subsubsection*{Acknowledgements}
The author, Raymond and Beverly Sackler Foundation Research Fellow at the IoA, 
thanks the Foundation for generous financial support, ESO for support at the 
meeting, and Kate Quirk for helpful comments on the manuscript.

\clearpage
\addcontentsline{toc}{section}{Index}
\flushbottom
\printindex

\end{document}